# First studies of electron transport along small gas gaps of novel foil radiation converters for fast-neutron detectors

M. Cortesi,[a,b,*] R. Zboray,[a] R. Adams,[a,b] V. Dangendorf,[c] A. Breskin,[d] S. Mayer,[a] H. Hoedlmoser[a] and H.-M. Prasser[b]

[a] *Paul Scherrer Institut (PSI), Villigen PSI, CH-5234, Switzerland*
[b] *Swiss Federal Institute of Technology (ETH), Zurich, CH-8092, Switzerland*
[c] *Physikalisch-Technische Bundesanstalt (PTB), Braunschweig 38116, Germany*
[d] *Weizmann Institute of Science, Rehovot, 76100, Israel*
  *E-mail*: marco.cortesi@psi.ch

ABSTRACT: Novel high-efficiency fast-neutron detectors were suggested for fan-beam tomography applications. They combine multi-layer polymer converters in gas medium, coupled to thick gaseous electron multipliers (THGEM). Neutron-induced scattering on the converter's hydrogen nuclei results in gas ionization by the escaping recoil-protons between two successive converters. The electrons drift under the action of a homogeneous electric field, parallel to the converter-foil surfaces, towards a position-sensitive THGEM multiplying element.

  In this work we discuss the results of a systematic study of the electron transport inside a narrow gap between successive converter foils, which affects the performance of the detector, both in terms of detection efficiency and localization properties. The efficiency of transporting ionization electrons was measured along a 0.6 mm wide gas gap in 6 and 10 mm wide polymer converters. Computer simulations provided conceptual understanding of the observations. For drift lengths of 6 mm, electrons were efficiently transported along the narrow gas gap with minimal diffusion-induced losses; an average collection efficiency of 95% was achieved for ionization electrons induced by few keV photoelectrons. The 10 mm height converter yielded considerably lower efficiency due to electrical and mechanical flaws of the converter foils. The results indicate that detection efficiencies of ~7% can be expected for 2.5 MeV neutrons with 300-foils converters, of 6 mm height, 0.4 mm thick foils and 0.6 mm gas gap.

KEYWORDS: Micropattern gaseous detectors; Neutron detectors; Neutron radiography.

---

[*] Corresponding author.

# Contents

## 1. Introduction

A detector for fast-neutron imaging has been developed for fan-beam neutron-tomography applications, as described in reference [1]. It combines (Figure 1): 1) a multi-layer neutron foil-converter made of hydrogenous non-scintillating polymer material (e.g. polyethylene), with large surface-to-volume ratio, converting neutrons into recoil protons; 2) surrounding gas, in which the neutron-induced recoil protons escaping from the converter foil release ionization electrons; 3) an electric field set parallel to the converter foils, to efficiently drift the ionization electrons along the gas gap between two successive converter foils; 4) a position-sensitive gas-avalanche imaging electron multiplier, with a dedicated data-acquisition system, to localize neutron's interaction location.

    The operation principle is the following: fast neutrons collide elastically with hydrogen nuclei (n-p scattering) in one of the multiple converter foils along the neutron's flight path. Scattered protons with sufficient kinetic energy can escape the foil and ionize the gas in the narrow gap between two successive foils. The ionization electrons are drifted by an electric field towards a gas avalanche electron multiplier, where they are detected and localized. The drift field, parallel to the converter foils, is maintained by a potential difference between the top and bottom of the converter.

    The critical issue for such a configuration is the electron drift along the narrow (a fraction of a millimeter) gas gap. Due to the transverse electron diffusion during its drift, and field distortions by material charging-up, some of the electrons may impact the foil surface; as a consequence, their collection efficiency by the multiplier may decrease with the drift distance, affecting the overall neutron detection efficiency.

    In this work we present a systematic study of properties and phenomena related to the transport of ionization electrons within the converter's small gas gaps. We have carried out

computer simulations and experimental investigations with converters produced as prototypes by industry. To avoid charging-up of the polymer foils and to maintain a constant undistorted field within the gas gaps, the converters were produced using relatively low-resistivity (anti-static/dissipative) materials (in this work antistatic Acrylonitrile Butadiene Styrene – ABS-EDS7 [2]). Also other materials were considered, for example. antistatic-Polyethylene – UHMW-PE [3], which has a better H/C ratio, but so far no manufacturer could be found. The converters were coupled to a position-sensitive Thick Gas Electron Multiplier (THGEM) [4], filled with Ne/5%CH$_4$ or Ne/5%CF$_4$ gas mixtures at atmospheric pressure (Figure 1); the operation of THGEM with Ne-based mixtures is attractive since it provides high electron gain at comparably low operational voltages. This is important to limit the energy dissipated in sparks (which is proportional to the square of the amplification voltage). Occasional sparking is difficult to avoid due to the high dynamic range of the primary ionization in the gas. For details about THGEM-based imaging detectors and their applications the reader is referred to [5,6] and references therein.

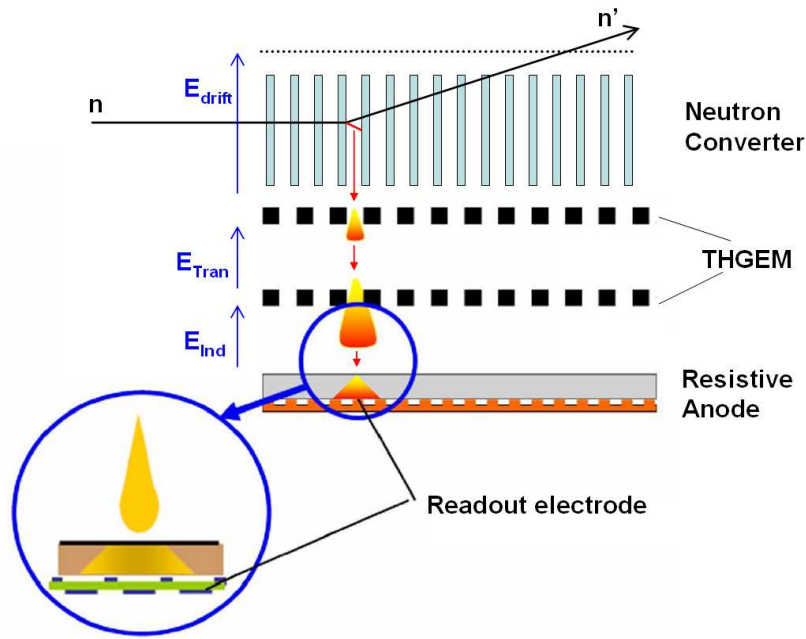

**Figure 1. Schematic drawing of the fast-neutron imaging detector concept.**

The electron-transport properties within the small gas gaps between converter foils and related phenomena (i.e. charging up effects, electron diffusion etc.) are independent of the primary particle which injected the primary charge into the gas. Due to the difficulties of handling neutron sources or generators in the laboratory (radiation protection issues, requirement of narrow, collimated pencil beams etc.) this investigations were carried out in this work with 5.9 keV X-rays irradiation (from a $^{55}$Fe source), depositing ~165 ionization electrons per event in the gas by photoelectric conversion. The experimental results from converter prototypes coupled to a 10x10 cm$^2$ THGEM imaging detector were compared to that of Monte-Carlo simulations.

We describe below the converters, experimental setups and methodology and report on the results obtained so far. The applicability of the new concept to fast-neutron detection and imaging, and the expected performance, are discussed.

## 2. Converter Prototypes

The optimal geometry of the converter prototypes, in particular the foil thickness and the gas gap width between two consecutive foils, were defined and calculated in a previous Monte Carlo study; details can be found in [1]. For the highest neutron-to-proton conversion efficiency, it was found that in the case of 2.5 MeV neutrons beam, the minimum size of the converter foils should be of ~0.1 mm (corresponding to the range of the most energetic recoil proton in polyethylene [1]). However, such converter structures, having long and thin converter foils, are mechanically fragile and difficult to manufacture with good accuracy. As a compromise, a larger foil thickness (e.g. 0.4 mm) has therefore to be considered, at the expense of somewhat smaller neutron detection efficiency [1]. As this work was mainly focused on understanding electron transfer properties along the small gas gap between the converter foils, the converter prototypes investigated here were not optimized for the highest neutron detection efficiency. In addition, although polyethylene (PE) is the preferable fast-neutron converter material with the highest hydrogen density, our present prototypes are made of ABS-ESD7 which are of simple and inexpensive production; ABS-ESD7 has the same density as PE (~1 $g/cm^3$) but about 20% less hydrogen content (namely expected ~20% lower neutron-conversion efficiency).

We have tested two converter prototypes (Figure 2) having different heights (10 mm and 6 mm along the Z coordinate); both cover an area of 10x10 $cm^2$ (X-Y coordinates); they are comprised of 83 foils of 0.6 mm thickness, displaced perpendicularly to the radiation direction, with a pitch of 1.2 mm; the resulting gas gap between consecutive layers is 0.6 mm. All converters were manufactured using 3D printing technology (3D printers use "additive fabrication" methods building polymer structures, layer by layer, of any geometrical shape).

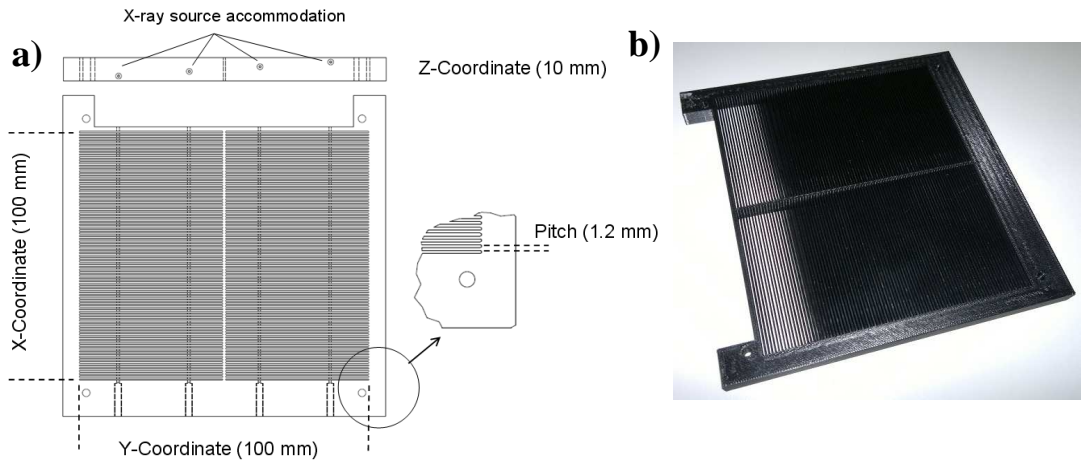

**Figure 2. Part a: fast-neutron converter prototype with four irradiation apertures for electron transport studies with X-rays at different drift lengths (for detailed description see the text). Part b: photo of the 6 mm ABS converter.**

Figure 2a shows a technical drawing of the 10 mm thick prototype; this particular converter structure has 4 apertures of 1 mm diameter drilled all along the X-axis for X-ray injection at four different heights (0.2 mm, 0.4 mm, 0.6 mm and 0.8 mm) along the Z-axis. They permit investigating radiation-induced electron transport for these four drift distances along the gas gaps.

Note that, according to the producer the bulk resistivity is in the range of $10^8$-$10^9$ Ohm·cm. However, due to the peculiar 3D layer-by-layer printing, the resistivity value is not fully isotropic; in particular, the 10 mm thick ABS converter is dissipative/antistatic (low resistivity) only along a single horizontal layer (X-Y plane in Figure 2) while the resistivity between the different layers is higher (direct measurements performed in the lab have yielded a bulk-resistivity values above $10^{13}$ Ohm·cm along the vertical direction – Z-axis in Figure 2). While the quest for better production methodologies and materials is ongoing, the systematic study, reported here, aims at characterizing the transport of electrons within the present rudimentary "resistive" converter prototypes.

## 3. Simulations of X-ray interaction in gas

The interaction of a 5.9 keV photon with a Ne atom results mainly in the emission of a primary photoelectron with an energy of 5 keV and an Auger electron of 0.9 keV (for Ne, the low probability X-ray fluorescence can be neglected).

The total number of electron-ion pairs deposited by these X-ray induced electrons in a Ne-based mixture (low percentage of quencher) can be expressed as

$$n_T = \frac{\Delta E}{W_i} \qquad \text{Eq. 1}$$

where $\Delta E$ is the total energy loss in the gas volume considered, and $W_i$ is the effective average energy to produce one pair. For Ne, $W_i$ = 36 eV [7]; as a result, neglecting effects of small $CH_4$ or $CF_4$ additives, a maximum number of ~165 electrons are released in our Ne mixtures by a 5.9 keV photoelectric absorption: these ionization electrons are mostly released along the photoelectron track. This maximum number is only achieved for those photoelectrons which are fully slowed down in the gas gap only and not hitting one of the foil surfaces.

Figure 3 depicts the distribution of deposited energy of 5 keV photoelectrons and 0.9 keV Auger electrons in Ne, at standard conditions of temperature and pressure; it was calculated using MCNP simulation package [8]. The computed value of the practical range is of the order of 0.9 mm for 5 keV photoelectrons, namely larger then the prototype's gas gap width between successive converter foils (0.6 mm). Therefore, only part of the 5 keV photoelectron's energy is deposited in the gas and converted into ionization electrons, depending on the position of the photoelectric interaction and the path of the emitted photoelectron (see Figure 4).

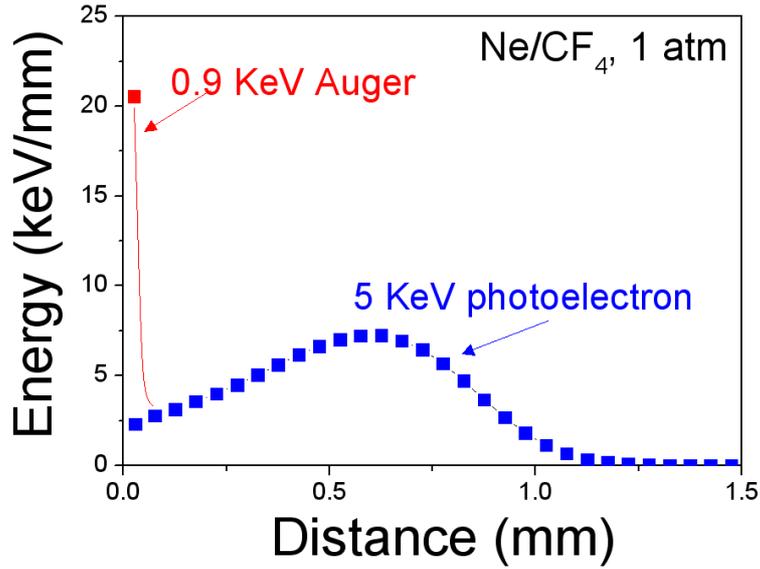

**Figure 3. Distributions of deposited energy in the Ne-based mixture by 0.9 and 5 keV x-ray induced electrons. The practical range of the 5 keV photoelectrons extends beyond the 0.6 mm gas gap width.**

Figure 4 is a snapshot from MCNP illustrating tracks of 5 keV electrons (100 simulated tracks). Two cases are shown: a) the electrons start form the center of the gas gap and b) the electrons start close to the converter foil (no electric fields were applied here).

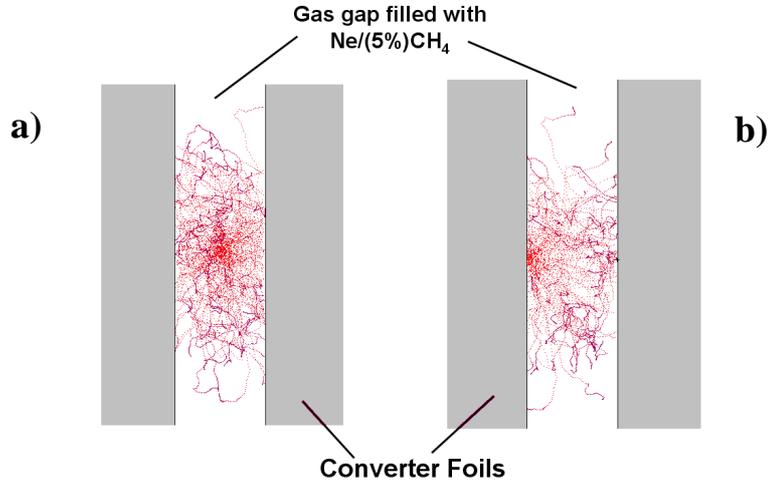

**Figure 4. MCNP snapshot from a side view of the converter, showing 5 keV simulated photoelectron tracks (100 tracks) in the 0.6 mm gaps filled with a Ne-mixture; a) tracks starting in the middle; b) tracks starting close to the foil. No electric drift-field was applied.**

### 4. Simulation of the experimental detector setup

Figure 5 depicts a schematic drawing of the detector setup and the irradiation condition used in our measurements. In the present detector configuration, the pitch of converter foils (1.2 mm) is larger than the pitch of the THGEM hole (1 mm), the latter is arranged in a hexagonal pattern.

In order to optimize the focusing of the electrons from this non-matching converter gas gap geometry into the THGEM holes, the converter is separated from the top electrode of the double-THGEM multiplier by a 3.2 mm long drift gap.

The THGEM geometry was the following: 0.4 mm thickness, 0.4 mm holes diameter, 0.1 mm etched holes rim, and 1 mm holes pitch. The double-THGEM operated at a total gain of ~$10^4$. Electric fields across the converter (X-Y direction in Figure 2) were maintained by means of fine electroformed meshes (93% transparent) on its top and bottom sides.

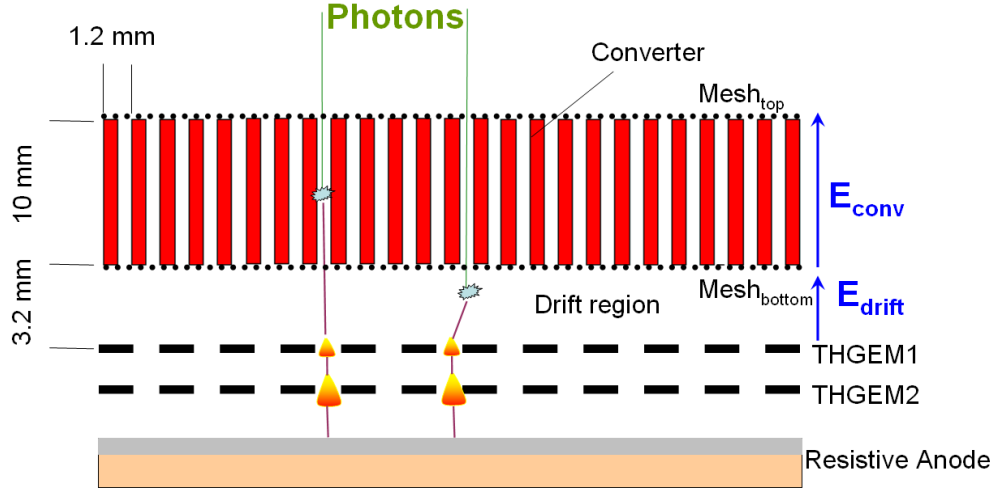

**Figure 5. Schematic drawing of the detector setup: a 3.2 mm thick drift region (stick to figure inscriptions) is maintained between the converter and the top THGEM electrode.**

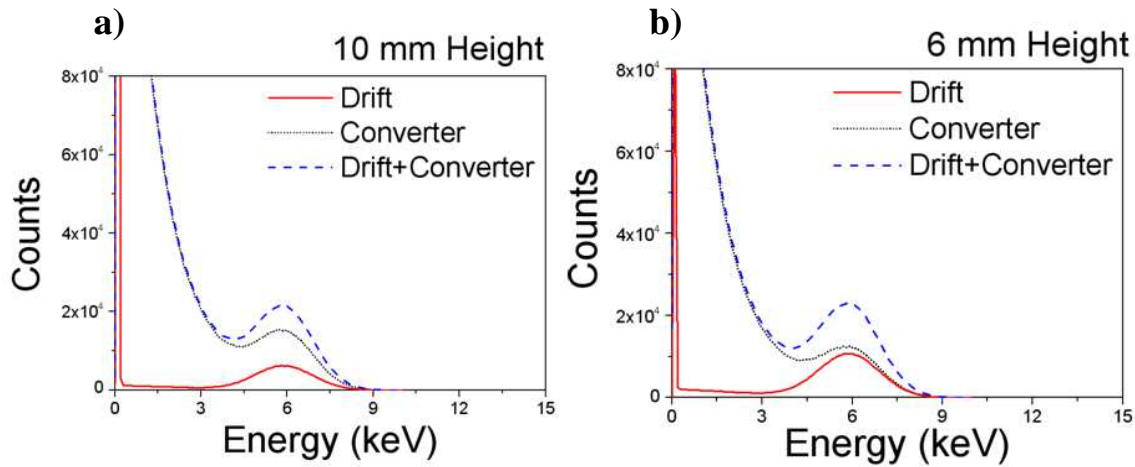

**Figure 6. Simulated energy distributions as detected in the detector setup shown in Figure 5, induced by 5.9 keV X-ray photons, with a) 10 mm high converter and b) 6 mm high converter. The red curve: charge collected from the 3.2 mm thick drift region; black curve: charge collected from the converters' gas gaps; blue curve: the sum of the red and black spectra. The distributions were convolved with the estimated THGEM response function.**

In this configuration the detector was irradiated from the top through a 1 mm diameter collimator. The X-ray photons were absorbed either in the converter's gas gaps or in the drift gap between the bottom mesh and the THGEM (see Figure 5). By setting adequate electric field values (and directions) across the converter and the drift gap, one could collect and detect the ionization electrons created in both regions; e.g., a reverse electric field across the converter permitted collecting electrons only from the drift gap.

Figure 6 shows the results of Monte Carlo simulations (MCNPX) modeling the detector shown in Figure 5. The energy distributions correspond to the yields measured in the detector with 6 mm (a) and 10 mm (b) high converters; they show spectra for photons absorbed in the converter gas (black curve), in the drift gap (red curve) and the sum of the two (blue curve). To represent a realistic detector response and for future comparison with experimental data, the spectra were convolved with the pulse-height response of a THGEM-based detector; we employed here a moderate value of 40% FWHM, resulting in rather broad distributions peaked at 5.9 keV (Figure 6). X-rays absorbed in the drift gap, with the full deposited energy transferred to the THGEM, yielded a narrow distribution peaked at 5.9 keV; the low-energy tail is due to partial energy deposition in the gas of many photoelectrons produced in the converter region.

Photoelectrons (and Auger electrons) released in the converter's gas gap, are absorbed in the gas or stopped at the foils; in the more frequent latter case, only part of their energy is deposited in the gas – resulting in the exponentially decreasing low-energy tails shown in Figure 6. The distributions in Figure 6, resulting from the converters of different heights, are qualitatively similar; but due to geometrical reasons the converter-to-drift gap ratio for the absorbed photons is smaller in the shorter converter.

## 5. Experimental results

### 5.1 Side-irradiation of the converter

Side-irradiation of the 10 mm thick converter with X-rays, through the small apertures (1 mm diameter) at various drift distances shown in Fig. 2, provided an experimental means for optimizing the field across the converter structure. It also permitted investigating the efficiency of the transport of ionization electrons across the small gas gaps, as a function of the drift length.

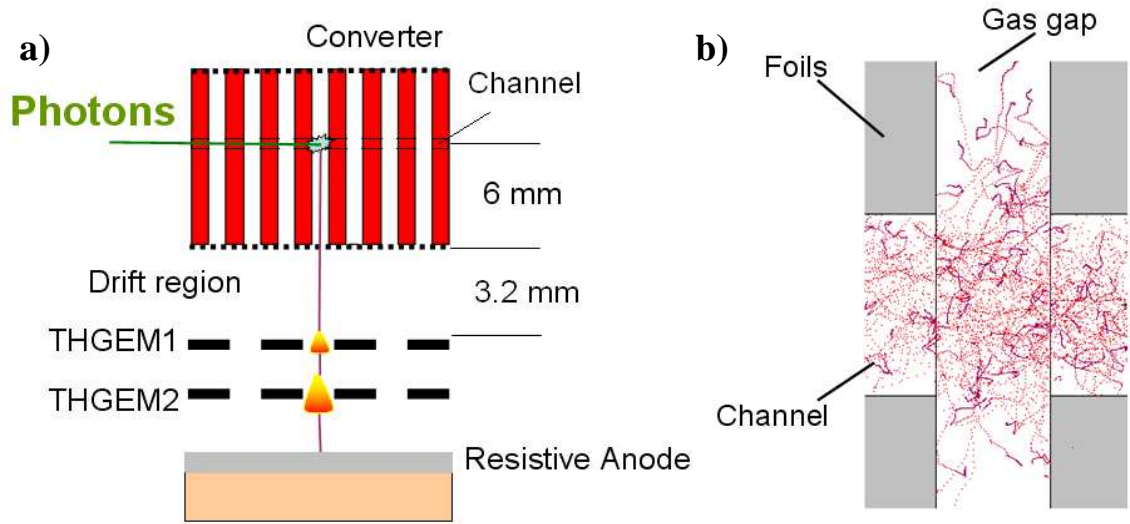

**Figure 7. (a) Schematic drawing of the side-irradiation geometry with the source placed 6 mm from the drift gap boundary. (b) Simulated snapshot of photoelectrons' tracks in the converter channel and in the gas gap; MCNP visualization tool (VISED).**

In this configuration (Figure 7a), X-ray photons can be absorbed along the drift channels, both in the gas gap between the foils as well as within the collimation hole in the converter foils. The emitted photoelectrons have a larger range than the width of the gap between two foils (see section 3), therefore they deposit energy inside the hole of the converter foil (ionization produced here is mainly lost for the drift process) and the gap between two adjacent foils (theses ionization electrons are relevant for the detection process). The share between these two depends on their interaction location and the details of the photoelectron trajectory (Figure 7b).

To estimate the number of relevant ionization electrons the energy deposition by the photoelectron was simulated with MCNP for a large number of X-ray histories. Figure 8 shows the simulated spectrum of energy deposited in the gas gap between two foils, resulting from X-rays absorbed along the collimation channel (Figure 7a). The spectrum was convolved with the detector response function (e.g. 40% FWHM). The spectrum of Figure 8 does not have the typical photo-absorption peak at 5.9 keV; the rather flat distribution results from the shared energy deposition in the gas gap and inside the hole of the foils.

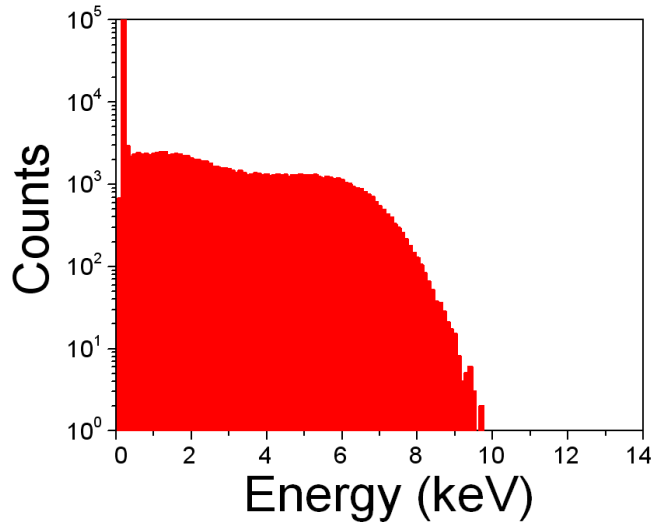

**Figure 8. Simulated spectrum of energy deposited in the gas gap, resulting from X-ray-induced photoelectrons along the collimation channel.**

The 10 mm thick converter was irradiated at 2 mm, 4 mm, 6 mm and 8 mm drift distances. Spectra were recorded at the different drift distances and as function of the field across the converter. Figure 9 depicts spectra measured for a 6 mm drift distance, for different electric fields applied across the converter. All spectra have exponentially decreasing distributions, with intensities varying with the electric field strength; the maximum electron collection efficiency was achieved for field values above 0.4 kV/cm. In all the aforementioned measurements the field across the drift gap was kept at 1 kV/cm.

X-ray induced spectra measured at drift distances of 2-8 mm are shown in Figure 10; the fields across the converter and the drift gap were kept at the values of 0.5 and 1 kV/cm, respectively. The spectra were fitted with an exponentially decreasing function and the corresponding areas under the curves were calculated. For longer drift lengths a significant loss of charge, mostly due to diffusion, was observed; compared to the 2 mm drift length, the spectra recorded at 4 mm and 6 mm have roughly the same loss of charge of around ~20%, while the 8 mm drift the spectrum is characterized by a loss of charge above 40%. Note that the shapes of the distributions are somewhat different compared to the ones computed with Monte Carlo simulations (Figure 8). The reason of this is because our MCNP computation code does not take into consideration electron transport, including electron diffusion, the real geometry of the field lines within the gas gap and other secondary effects; among them the possible charging up of the polymer foils that may affect the detected pulse- height distribution.

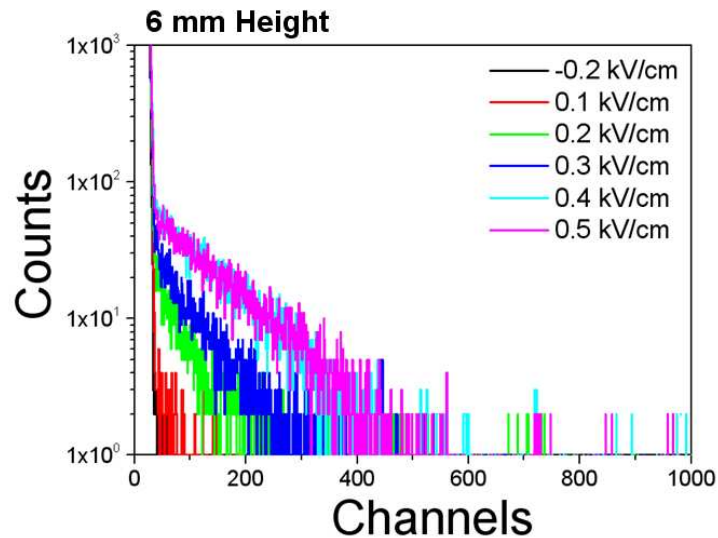

**Figure 9. X-ray induced spectra measured while irradiating the detector of Figure 7a through the collimation channel located at 6 mm above the drift gap; the drift field values across the converter are indicated. Field across the drift gap: 1kV/cm. All measurements performed at the various drift length channels show a similar saturation curves, though with different transport efficiency.**

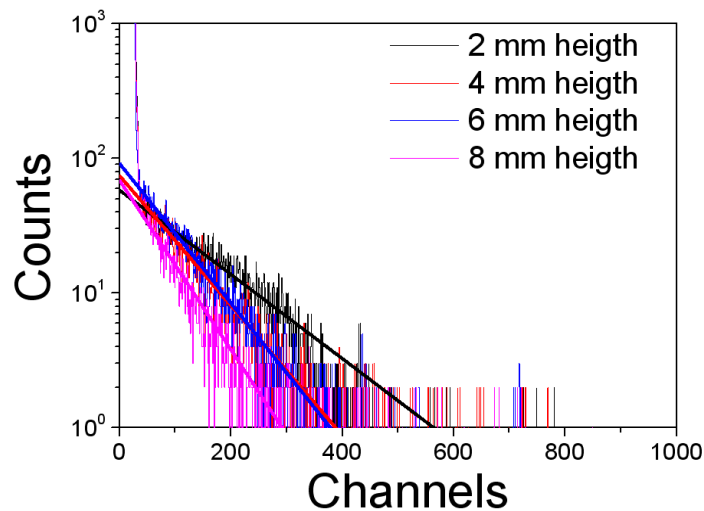

**Figure 10. X-ray induced spectra measured in detector depicted in Figure 7a, irradiating through the collimation channel located at different heights above the drift gap, corresponding to drift lengths of 2-8 mm inside the converter stack. Electric fields of 0.5 kV/cm and 1 kV/cm were kept across the converter gas gaps and the drift gap respectively. Same irradiation times for all conditions.**

**5.2 Top irradiation: X-ray spectra**

Figure 11 depicts X-ray spectra obtained with the detector configuration illustrated in Figure 5, with both 10 mm and 6 mm high converters; two different electric-field strengths were set across the converters: a "reverse" drift field of –0.4 kV/cm (red curves) and a "forward" 0.5 kV/cm field (blue curves). The electric field strength across the drift gap was kept at 1 kV/cm. Measurements with both converters were performed under similar operation and irradiation conditions, with the X-ray source was placed above the converter and the spectra recorded for 1000 s irradiation times. The pulse height spectra obtained with the 10 mm converter have a lower counting statistics compared to the one with the 6 mm converter. However, relevant for the comparison with the simulated electron transport efficiencies is the relation between spectra measured with reverse and forward electric fields across the converter gas gap. Within the measurements of each configuration (6 and 10 mm height converter) these spectra were obtained using the same number if incident photons. Charge signals for pulse-height analysis were recorded from the top electrode of the second THGEM element; they were processed by a charge sensitive preamplifier (ORTEC 142A), a linear shaping amplifier (ORTEC 572) and a multichannel analyzer (MCA).

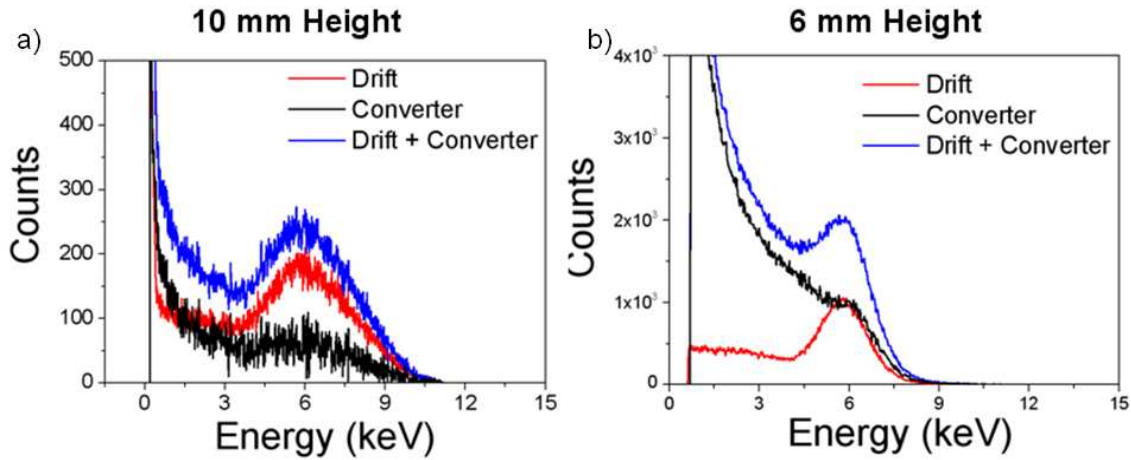

**Figure 11. X-ray charge spectra recorded with the detector of Figure 5; a) data with a 10 mm thick converter; b) data with a 6 mm thick converter. The red curves were obtained with a reverse (–0.4 kV/cm) field across the converter; the blue curves were recorded with a forward collection-field of 0.5 kV/cm across the converter. In both cases, the electric field across the drift gap was of 1 kV/cm. The black graph (subtraction of the two spectra) represents converter events only.**

The red curves of Figure 11 were obtained with a reverse electric field across the converter; in this configuration, all ionization electrons deposited within the converter gas were drifted away and collected by the top mesh electrode; thus the data represent only the ionization events deposited in the drift region.

The blue curves (drift+converter events) in Figure 11 depict spectra obtained under forward collection fields. Like in the simulations, the distributions are peaked at 5.9 keV, preceded by low-energy tail due to X-ray photons interacting in the gas gap of the converter. The difference between the integral of the blue and the red curves in Figure 11 thus corresponds to spectra of the charges collected from the converter.

The comparison between the simulated and experimental spectra offers a way to estimate the efficiency of collecting the drifting electrons from the thin gas gaps of the converter and the efficiency to detect the original radiation events. A deviation from the simulated spectra (where all the deposited energy in the gas is detected) would indeed be an indication of charge losses during the electrons drift inside the thin gas gaps, due to diffusion, charging up or other effects related to the transport of the electrons, from their creation point inside the converter to the THGEM multiplier.

In particular, let us consider the number of events in the peak areas (after Gaussian fitting and integration – see Figure 12), calculated in the drift+converter and in the drift gap spectra obtained with the Monte Carlo computation (Figure 6). The ratio of these two numbers is of the order of 3.8:1 and 2.4:1 for the 10 and 6 mm converters, respectively; this is the value of reference, corresponding to the ideal case of a perfect detector, with full detection efficiency and no loss of charges.

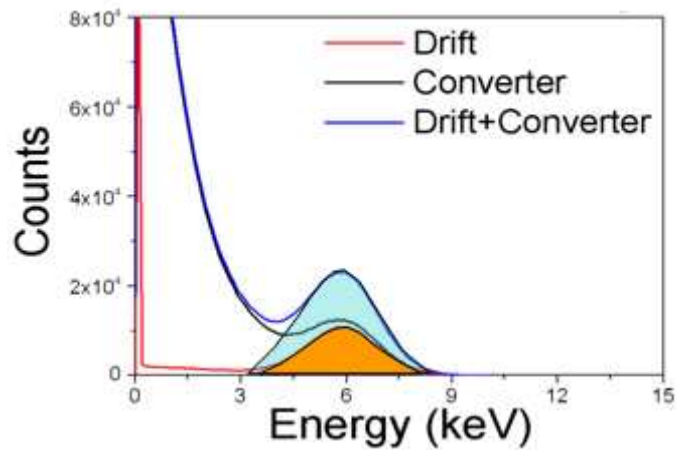

**Figure 12. Details of the Gaussian fitting of the photoelectron peaks (blue for the Drift+Converter spectrum and Orange for the Drift spectrum) for the spectra of deposited energy calculated using MCNP.**

As shown in Figure 11a, the experimental data with the 10 mm thick converter display a significant discrepancy with the corresponding simulated spectrum (Figure 6a). The measured ratio 1.34:1 of the counts in the converter and in the drift region under the peak is very low compared to the simulated one. The efficiency of detecting the full energy of photoelectron conversion (5.9 keV peak) is thus of the order of only 30%, indicating a serious distortion of the electric field geometry within the small gas gaps. As mentioned before the 10 mm converter showed strong non-isotropic conductivity, in particular in drift direction. Therefore the loss of ionization electrons is most probable caused by electron attachment and successive up-charging of the converter walls. On the other side the 6 mm thick converter displays larger electron transport efficiency; in the peak region the ratio of number of drift+converter electrons to the number of drift gap electrons is of 2.2:1, , which is in good agreement with the Monte Carlo value (Figure 6b). In this case the efficiency of detecting the full photoelectron-induced electrons is of the order of 92%; computer simulations proved that the loss of efficiency for 6 mm is due to electron diffusion (see section 5.4).

In Figure 13 experimental and simulated spectra corresponding to photons absorbed in the drift region (a) and in the converter region (b) are compared; the data were taken from Figures 6b (simulation results) and Figure 11 (experimental data). The MCNP results were rescaled such that the two spectra have the same number of counts in the peak region. There is a good qualitative agreement between the experimental and simulated spectra, though in the case of drift gap absorption, (a), the simulated data underestimate the low-energy tail by a factor of 2-3. This could result from the fact that the simulations do not take into consideration the exact field geometry (some field lines in the drift gap do not guide electrons into a hole but end on the THGEM top surface where an electron would be attached and therefore is lost for detection) and electron diffusion - both causing losses of ionization electrons and therefore of efficiency. Better agreement between experimental and simulation data is shown in the converter-absorption graph (b), despite a significantly poorer energy resolution which smoothens the peak.

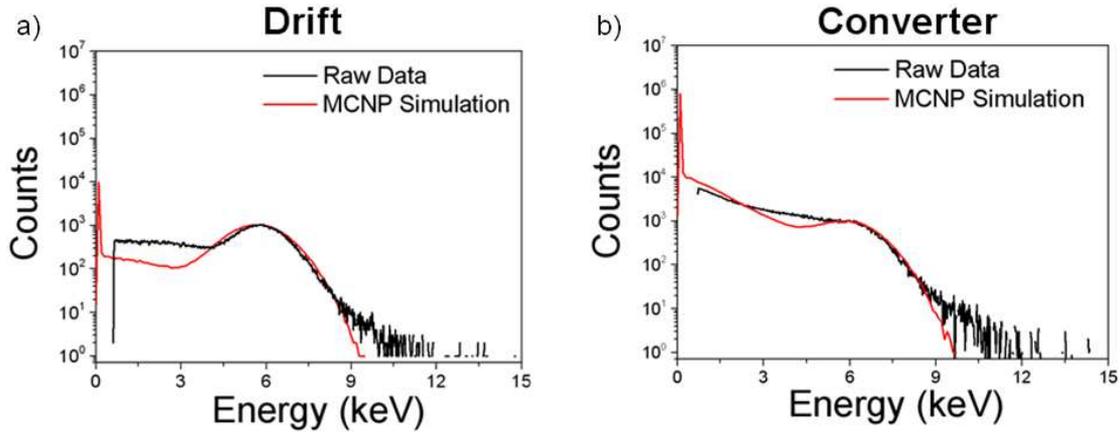

**Figure 13. Comparison between the experimental spectrum (black) obtained with the 6 mm height converter and the simulated deposited energy spectrum (red curve); a) photons absorbed only in the drift gap; b) photons absorbed only in the converter.**

### 5.3 Tuning the electric fields

When the converter is separated from the first THGEM by a small drift gap, it is necessary to optimize the ratio of the field in the drift gap to the one across the converter in order to maximize the collection of the ionization electrons from the converter gas gaps into the drift gap, and transfer them further towards the THGEM multiplier.

Figure 14 illustrates different field geometries computed by Maxwell SV (ANSYS) [9]. Notice that, for sake of simplicity, in these simulations the detector setup was designed such that the THGEM holes' pitch was matching the converter foils pitch, differently from the detector configuration tested in this work; however, this geometrical discrepancy is not relevant within the context of the arguments discussed here. In addition, in order to assure a homogenous electric field inside the gap pointing towards the direction of the THGEM, the foils were subdivided in discrete layers of 0.1 mm width, each biased at a different potential.

At the interface between converter and drift gap, the field distribution depends on the ratio between the respective electric field strengths; it affects electron defocusing or focusing during their transport from the converter towards the THGEM. Figure 14 show three different field ratios, 2:1, 1:1 and 1:2 respectively; the drift-gap field was kept at 1 kV/cm. For field ratio smaller than 1:1, the electrons are defocused at the interface between the converter and the drift gap, causing a significant reduction of the charge transport towards the THGEM multiplier.

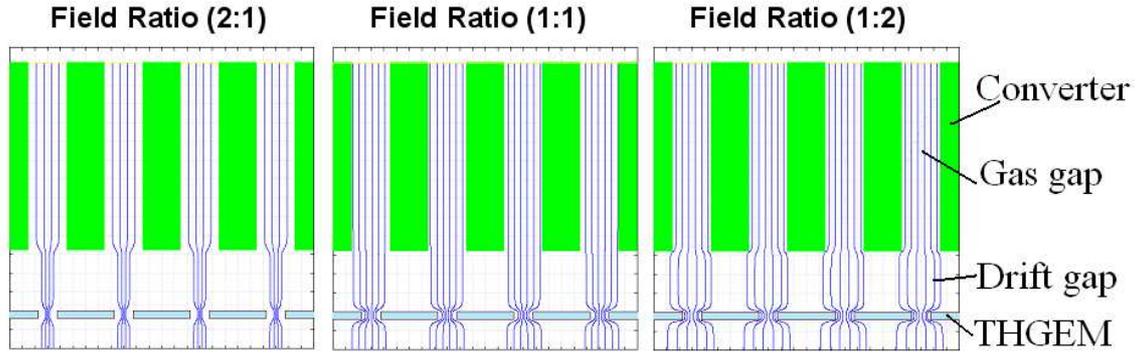

**Figure 14. Effect of electric-field focusing at the drift/converter interface as a function of the drift-gap to converter electric field ratio (values shown in the figure).**

The loss of charges due to field defocusing at the converter/drift-gap interface is also evident from the experimental $^{55}$Fe raw spectrum (black graph) depicted in Figure 15; it was recorded with fields of 1.5 kV/cm and 1 kV/cm applied to the 10 mm thick converter and the 3.6 mm thick drift gap, respectively. The experimental spectrum (purple graph in Figure 15) was reconstructed using the simulation results of Figure 6 (in Figure 15, red and blue curve respectively for the converter and drift-gap simulation curves). The fitting was analytically performed using least square with MATLAB [10]. The process consisted of minimizing the residual between experimental data $CD_{exp}$ and the sum of the drift $D_{MC}$ and the converter curve $C_{MC}$ from simulation, by tuning the amplitude of the latter ('$a$' and '$c$') as well as rescaling it along the energy axis ('$b$' in Eq. 2).

$$\min_{a,b,c} \left[ CD_{exp}(E) - (a*D_{MC}(E) + c*C_{MC}(b*E)) \right]^2 \qquad \text{Eq. 2}$$

The significant shift towards the low-energy region of the converter spectrum peak (red graph) compared to the drift-gap spectrum peak (blue graph), is a clear indication of losses of charge at the interface between the converter and the drift-gap due to field defocusing. Through a systematic investigation of the transport of charge at the interface between the two, trying different combinations of voltages, it was found that electrons are efficiently transferred and focused into the THGEM holes for field ratios larger than 2:1.

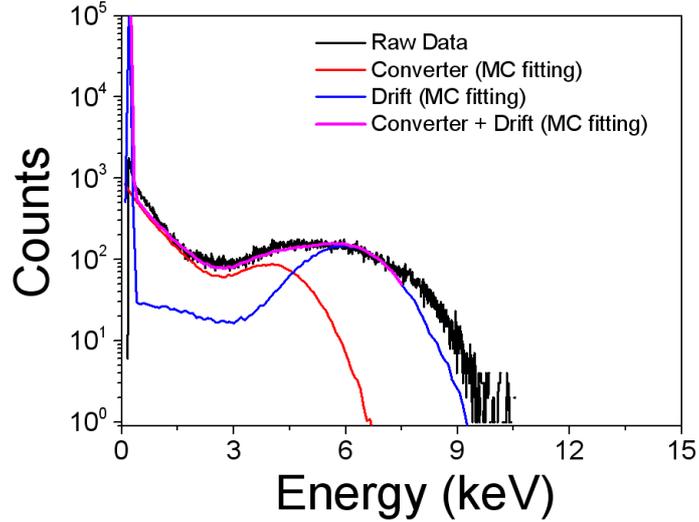

**Figure 15.** Reconstruction of the experimental spectrum measured (detector of Figure 7) with 5.9 keV X-rays, using the simulation spectra of Figure 6. The electric field values across the converter and the drift gap were 1.5 kV/cm and 1 kV/cm, respectively.

### 5.4 Electron transport in the gas

Electrons hitting the surface of the converter foils are lost for the drift process. The probability for electrons hitting the surface of the narrow gaps depends on the transverse charge distribution which is determined by diffusion of the drifting electron cloud. The diffusion and therefore width of the transverse charge distribution depends on the square root of the drift length. Both the drifting-electron's energy distribution and their diffusion depend on the electric field. As a consequence, in our geometry the probability of losing electrons increases with the drift distance.

We have carried out a systematic computer simulation study to evaluate the loss of charges due to electron diffusion as function of the drift length. Our simulations were performed using Garfield [11] for the computation of the electron transport properties in Ne-based gas mixtures; the 3D electric-field maps were computed by Maxwell SV.

For the computation of the electron transport efficiency we simulated 100 events, each generating 100 electrons randomly distributed in the gas gap along the plane connecting two foils; these electrons were drifted in an homogenous electric field for distances of 2-8 mm. The electric field along the converter gas gap was 0.5 kV/cm; the one along the drift gap was 1 kV/cm. The computed distributions of electrons reaching the THGEM detector and the resulting transport efficiencies are shown in Figure 16; the transport efficiency was defined as the percentage of events where at least one electron out of the 100 generated ones reached the THGEM element and was detected.

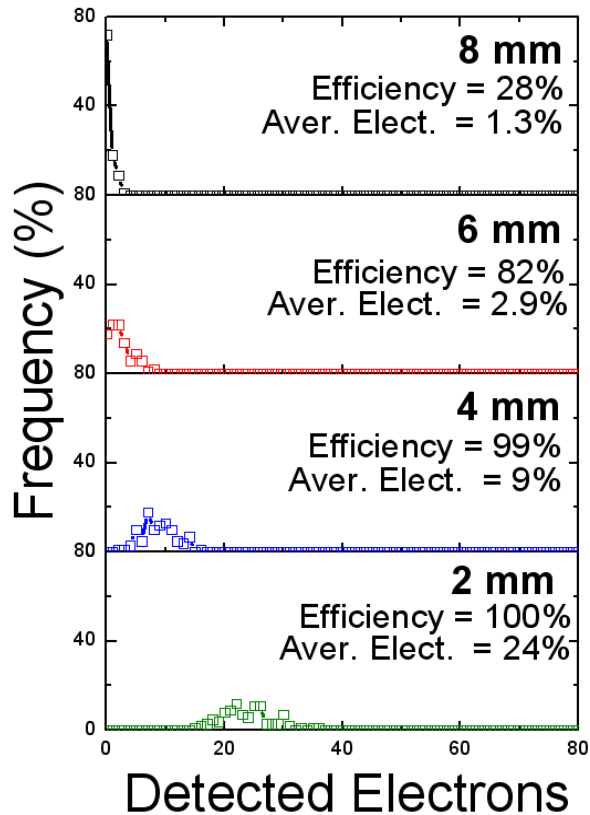

**Figure 16.** Simulated distributions of the fraction of the number of electrons (100 initial electrons generated in each event) reaching the detector, after drifting along different distances (indicated) within the 0.6 mm wide gas gap. The average numbers of arriving electrons and the event's detection efficiencies are indicated for each drift distance. The electric fields were 0.5 kV/cm across the converter and 1 kV/cm in the drift gap.

As shown in Figure 16, the electron transport efficiency, as well as the amount of electrons reaching the detector depends on the drift length. E.g., for a 2 mm drift, on the average 24% of the generated electrons reach the THGEM, each event resulting in at least one detectable electron reaching the THGEM. For larger drift lengths, the loss of charges increases, leading to significant decrease of the transport efficiency; e.g. for 8 mm drift length, on the average only 3% of the generated electrons reach the THGEM and the percentage of detectable events is only of 28%. Considering the converter with a 6 mm long gas gap (Table 1), the average transport efficiency estimated along the full converter length is about 95% (for 100 initially deposited electrons), which is very close the value observed experimentally (92%) irradiating the detector with 5.9 keV X-rays (~165 initially deposited electrons). In the case of 10 mm long converter, the expected average transport efficiency calculated along the full converter length is of the order of 70% (Table 1); the much lower value (30%) measured with the 10 mm converter prototype suggests that in this converter also other effects than just transverse diffusion play a role. Due to the lack of conductivity in drift direction in the 10 mm converter the electrons which attach to the foil surface are not neutralized but lead to up-charging. This modifies the field in-between the foils (e.g. screens the external field) and causes further losses of electrons.

**Table 1. Calculated and measured average electron transport efficiency for various converter heights,**

| Converter Height (mm) | Calculated average electron transport efficiency | Measured average electron transport efficiency |
|---|---|---|
| 2 | 100% | |
| 4 | 100% | |
| 6 | 95% | 92% |
| 8 | 80% | |
| 10 | 70% | 30% |

The simulation was repeated to estimate the fast-neutron detection efficiency with the present multi-foils converter. The number of electrons released in each event was defined according a realistic energy loss spectrum of 2.5 MeV neutron-induced recoil protons in the 0.6 mm gas gap. A typical spectrum of energy deposited by recoil protons scattered in the forward direction by is shown in Figure 17; it was computed by GEANT4 [12] and previously reported in [1].

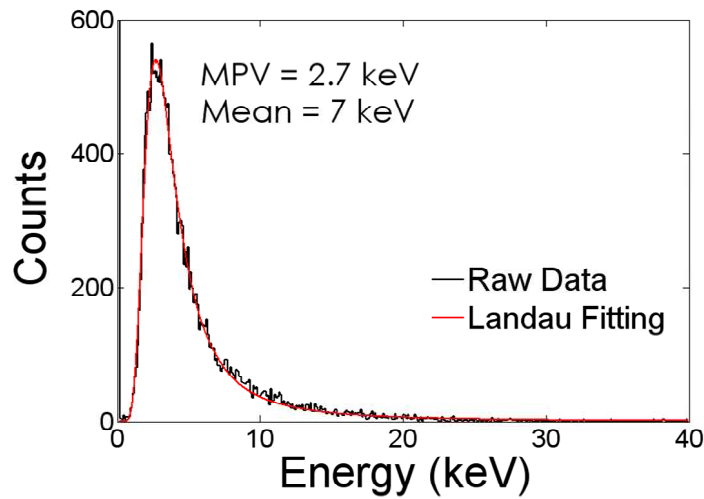

**Figure 17. Typical spectrum of energy deposited in 0.6 mm gas gaps by recoil protons scattered in forward direction by 2.5 MeV neutrons, computed with Monte Carlo simulation (black graph) and curve fitting with Landau distribution (red curve). Data taken from [1]. MPV stands for "most probable value".**

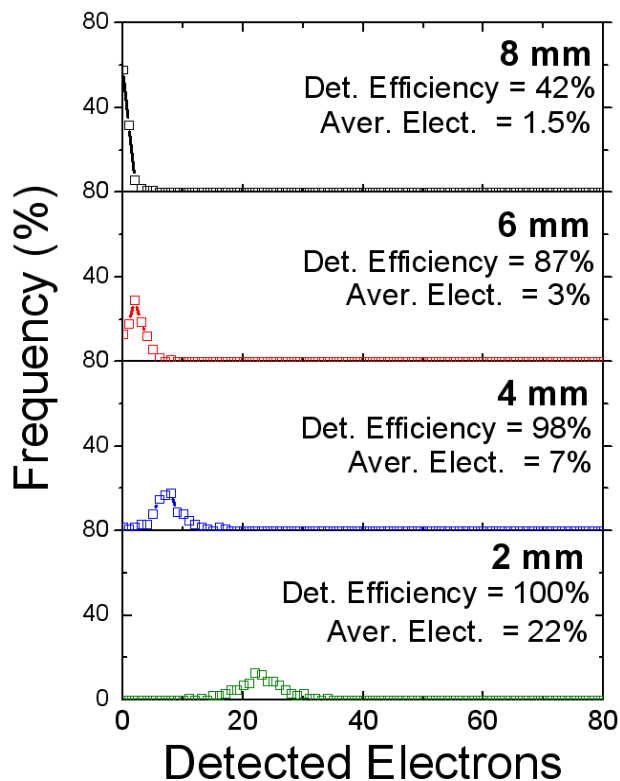

**Figure 18.** Distributions of the number of detected electrons and the respective electron transport efficiencies, simulated in the detector shown in Figure 5, after different drift distances. The number of deposited electrons, by forward-emitted protons resulting from 2.5MeV neutrons scattered on the converter foils, was derived from the energy distribution of Figure 17.

The energy loss spectrum of Figure 17 has the shape of a Landau distribution with a mean value of 7 keV – according Eq. 1 this means ~200 ionization electrons released on the average in the thin gas gap. The higher average number of initial ionization electrons, resulted in larger detection efficiencies (compared to Fig. 15) for the various drift lengths (see results shown in Figure 18). Even for 8 mm drift 42% of the neutron-induced events are detectable. The 6 mm high converter is expected to yield an average transport efficiency (at least one electron transfer to the charge detector) of the order of 97%. Note that in this analysis we did not consider any threshold discrimination of the THGEM detector signals, which eventually would lead to some loss of detection efficiency.

## 6. Discussion and Conclusion

The present work is part of a program aiming at the development of a novel fast-neutron detector concept for fan-beam tomography applications [1]. The detector is comprised of a series of hydrogenous (polymer) converter foils, arranged perpendicularly to the direction of the impinging neutron beam, coupled to a position-sensitive gas-avalanche detector for charge readout (here a double-THGEM, but other multiplier structures can be foreseen). This new detector concept offers the possibility for an efficient 1-dimensional spatial localization of fast-neutrons, with good spatial resolution due to the short range of the recoil proton within the small inter-foil gas gap.

In this work we have investigated properties and phenomena related to the transport of electrons within the small gas gaps - important for optimizing the collection efficiency of neutron-induced charges. The latter affects both the neutron detection efficiency and spatial resolution.

Transport of ionization charges through 6-10 mm high converter prototypes, with 0.6 mm wide gas gap, was investigated with soft X-rays. A series of Monte Carlo simulations (using GEANT4, MCNP and GARFIELD packages), modeling the detector and the irradiation configurations, permitted understanding the key physical processes underlying the experimental results.

It was found that for low drift lengths (foils of 6 mm height), loss of charge is expected mostly due to electron diffusion In this configuration a good electron collection efficiency, around 92%, was measured. For the measurement of 10 mm drift lengths with another converter the efficiency was significantly reduced and not in accordance with the expectations. This can be attributed to up-charging of the converter foils and successive screening and other distortions of the electric field in the drift region. The total detection efficiency (DE) of a potential fast-neutron imaging detector based on the present concept, can be written as the product of the neutron conversion efficiency $n_F$ (which depends of the number of the converter foils and their thickness), the electron transport efficiency $\varepsilon_T$ (depending on the drift length, namely the height of the converter foils) and an additional factor ($\varepsilon_{SP}$) which takes into account signal processing and electronic thresholds imposed by ionizing background and noise:

$$DE = n_F \, \varepsilon_T \, \varepsilon_{SP} \qquad \text{Eq. 3}$$

For a converter structure consisting of 300 foils, made of PE each of 0.4 mm thickness, it was calculated that the neutron-to-proton conversion efficiency ($n_F$) is of the order of 8% at 2.5 MeV [1]; the electron transfer efficiency ($\varepsilon_T$) for an axial resolution (converter height) of 6 mm was measured in this work to be of the order of 92%; assuming a THGEM-based detector gain of about $10^4$, a conservative signal-processing upper cut ($\varepsilon_{SP}$) of 10% could be applied [13]. Combined all these factors, for 2.5 MeV neutrons a total detection efficiency of 7% can be expected. Measurements of the gamma sensitivity are not presented here. Due to the long track and small energy deposition in a single gas gap, low gamma sensitivity can be foreseen; its value estimate is however out of this work's scope. This relatively high detection efficiency will be accompanied by mm-scale localization properties along the neutron beam direction, due to the neutron-induced protons stopping between foils. In addition, modern neutron sources can deliver practical rates in a tomographic facility of $\sim 10^7/s/cm^2$ (e.g. [14]). Given charge distribution among several converter foils and detection channels, the proposed THGEM-based detector can handle these rates [15] - provided that a suitable data acquisition system is employed.

In summary, our new concept is expected to offer advantages for 1D localization of fast-neutrons for a wide range of fan-beam tomography applications. Particular interest exists for non-destructive testing in the nuclear energy industry, such as multi-phase flow of optimization of fuel rod bundle geometry, spent nuclear fuel bundles inspection, safeguards. Other applications include detection of special nuclear materials (SNM), material science and detection of explosives (border control). The ongoing research program on this new detector concept includes convertor material studies and larger detector prototype investigations with neutron beams.

**Acknowledgments**

M. Cortesi gratefully acknowledges the support of the Swiss Society of Friends of the Weizmann Institute of Science. A. Breskin is the W.P. Reuther Professor of Research in the peaceful use of Atomic Energy. R. Adams is supported by the Swiss National Science Foundation (SNSF) grant number: 200021-12987C/1.